\documentclass[proceedings]{rmaa}

\newcommand{\ibid}{\rule[-0.03cm]{4.7em}{.01cm}.\ }

\def\plotBTD#1#2{%
  \expandafter\ifx\csname epsfbox\endcsname\relax
    \immediate\write16{%
        You need to input epsf; I'll do it for you
    }%
    \input epsf
  \fi
  \epsfysize=#2
     \openin 1 #1 \ifeof 1
        \immediate\write16{Can't open #1}%
        \vskip \the\epsfysize
      \else
         \closein 1
         \centerline{\epsfbox{#1}}%
      \fi
}

\title{Fast Reconnection of Magnetic Fields in Turbulent Fluids}

\author{A. Lazarian \affil{Department of Astronomy, University of
    Wisconsin-Madison, USA} \and
  E. Vishniac \affil{Department of Physics and Astronomy, Johns
    Hopkins University, Baltimore, USA}}

\fulladdresses{ \item A. Lazarian: Dept. of Astronomy, University of Wisconsin,
5534 Sterling Hall, 475 N. Charter St., Madison, WI 53706, USA
(lazarian@astro.wisc.edu)  
\item E. Vishniac: Dept. of Physics and Astronomy, Johns Hopkins University,
3400 N. Charles St., Baltimore, MD 21218, USA (ethan@pha.jhu.edu)}

\shortauthor{Lazarian \& Vishniac}
\shorttitle{Fast Turbulent Reconnection}

\resumen{La reconecci\'on es el proceso mediante el cual los campos
  magn\'eticos cambian de topolog\'{\i}a en un medio conductor y es
  fundamental para entender diferentes procesos, incluyendo la
  turbulencia interestelar y los dinamos estelares y gal\'acticos.
  Para explicar el campo gal\'actico y las r\'afagas y el ciclo solar,
  la reconecci\'on debe ser r\'apida y propagarse a la velocidad de
  Alfv\'en. Trabajos anteriores consideraban fluidos laminares y
  obten\'{\i}an tasas de reconecci\'on peque\~nas.  Mostramos que la
  presencia de una componente aleatoria del campo magn\'tico permite
  una {\it reconnecci\'on r\'apida} ya que, a diferencia del caso
  laminar donde el proceso avanza l\'{\i}nea por l\'{\i}nea, en el
  caso turbulento participan muchas l\'{\i}neas simult\'aneamente.  Una
  fracci\'on importante de la energ\'{\i}a magn\'etica se va a la
  turbulencia MHD, lo cual aumenta la tasa de reconecci\'on al
  aumentar la parte aleatoria del campo. Como consecuencia, los
  dinamos solares y gal\'acticos tambi\'en se vuelven r\'apidos.}

\abstract{Reconnection is the process by which magnetic fields in a
  conducting fluid change their topology. This process is essential
  for understanding a wide variety of astrophysical processes,
  including stellar and galactic dynamos and astrophysical turbulence.
  To account for solar flares, solar cycles and the structure of the
  galactic magnetic field reconnection must be fast, propagating with
  a speed close to the Alfv\'en speed.  Earlier attempts to deal with
  magnetic reconnection assumed that magnetized fluids are laminar and
  as a result obtained slow reconnection rates.  We show that the
  presence of a random magnetic field component substantially enhances
  the reconnection rate and enables {\it fast reconnection}, i.e.
  reconnection that does not depend on fluid resistivity. The
  enhancement of the reconnection rate is achieved via a combination
  of two effects.  First of all, only small segments of magnetic field
  lines are subject to direct Ohmic annihilation. Thus the fraction of
  magnetic energy that goes directly into fluid heating goes to zero
  as fluid resistivity vanishes. However, the most important
  enhancement comes from the fact that unlike the laminar fluid case
  where reconnection is constrained to proceed line by line, the
  presence of turbulence enables many magnetic field lines to enter
  the reconnection zone simultaneously. A significant fraction of
  magnetic energy goes into MHD turbulence and this enhances
  reconnection rates through an increase in the field stochasticity.
  In this way magnetic reconnection becomes fast when field
  stochasticity is accounted for. As a consequence solar and galactic
  dynamos are also fast, i.e. do not depend on fluid resistivity.}


\begin{document}

\maketitle
\vspace*{-4ex}
\section{Flux Freezing \& Reconnection}

Plasma conductivity is high for most astrophysical circumstances.
This suggests that ``flux freezing'', where magnetic field lines move
with the local fluid elements, is usually a good approximation within
astrophysical magnetohydrodynamics (MHD).  The coefficient of magnetic
field diffusivity in a fully ionized plasma is $\eta=c^2/(4\pi
\sigma)=10^{13}T^{3/2}$~s$^{-1}$ cm$^2$ s$^{-1}$, where $\sigma=10^7
T^{3/2}$ s$^{-1}$ is the plasma conductivity and $T$ is the electron
temperature. The characteristic time for field diffusion through a
plasma slab of size $y$ is $y^2/\eta$, which is large for any
``astrophysical'' $y$.

What happens when magnetic field lines intersect? Do they deform each
other and bounce back or they do change their topology? This is the
central question of the theory of magnetic reconnection.  In fact, the
whole dynamics of magnetized fluids and the back-reaction of the
magnetic field depends on the answer.

Magnetic reconnection is a long standing problem in theoretical MHD.
This problem is closely related to the hotly debated issue of the
magnetic dynamo (see Parker 1979; Moffatt 1978; Krause \& Radler
1980). Indeed, it is impossible to understand the amplification of
large scale magnetic fields without a knowledge of the mobility and
reconnection of magnetic fields.  Dynamo action invokes a constantly
changing magnetic field topology\footnote{Merely winding up a
  magnetic field can increase the magnetic field energy, but cannot
  increase the magnetic field flux. We understand dynamos in the
  latter sense.  The Zel'dovich ``fast'' dynamo (Vainshtein \&
  Zel'dovich 1972) also invokes reconnection for continuous dynamo
  action (Vainshtein 1970).} and this requires efficient reconnection
despite very slow Ohmic diffusion rates.

\section{The Sweet-Parker Scheme and its Modifications}

The literature on magnetic reconnection is rich and vast (see e.g.,
Biskamp 1993 and references therein). We start by discussing a robust
scheme proposed by Sweet and Parker (Parker 1957; Sweet 1958).  In
this scheme oppositely directed magnetic fields are brought into
contact over a region of $L_x$ size (see Fig.~1). The diffusion of
magnetic field takes place over the vertical scale $\Delta$ which is
related to the Ohmic diffusivity by $\eta\approx V_r \Delta$, where
$V_r$ is the velocity at which magnetic field lines can get into
contact with each other and reconnect. Given some fixed $\eta$ one may
naively hope to obtain fast reconnection by decreasing $\Delta$.
However, this is not possible. An additional constraint posed by mass
conservation must be satisfied. The plasma initially entrained on the
magnetic field lines must be removed from the reconnection zone. In
the Sweet-Parker scheme this means a bulk outflow through a layer with
a thickness of $\Delta$.  In other words, the entrained mass must be
ejected, i.e., $\rho V_r L_x = \rho' V_A \Delta$, where it is assumed
that the outflow occurs at the Alfv\'en velocity.  Ignoring the
effects of compressibility, then $\rho=\rho'$ and the resulting reconnection
velocity allowed by Ohmic diffusivity and the mass constraint is
$V_r\approx V_A {\cal R}_L^{-1/2}$, where ${\cal R}_L^{-1/2}=(\eta/V_A
L_x)^{1/2}$ is the Lundquist number.  This is a very large number in
astrophysical contexts (as large as $10^{20}$ for the
Galaxy) so that the Sweet-Parker reconnection rate is negligible.
\enlargethispage{2ex}
%\begin{figure}[h]
%\begin{center}
%\leavevmode
%\includegraphics[height=8cm]{vishniacfig1.ps}
%  \caption{({\it Top}) Sweet-Parker scheme of reconnection. ({\it
%      Middle}) The new scheme of reconnection that accounts for field
%    stochasticity. ({\it Bottom}) A blow up of the contact region. 
%Thick arrows depict outflows of plasma.
%}
%\end{center}
%\end{figure}

It is well known that using the Sweet-Parker reconnection rate it is
impossible to explain solar flares and it is impossible to reconcile
dynamo predictions with observations.  Are there any ways to increase
the reconnection rate?  In general, we can divide schemes for fast
reconnection into those which alter the microscopic resistivity,
broadening the current sheet, and those which change the global
geometry, thereby reducing $L_x$.  An example of the latter is the
suggestion by Petschek (1964) that reconnecting magnetic fields would
tend to form structures whose typical size in all directions is
determined by the resistivity (`X-point' reconnection).  This results
in a reconnection speed of order $V_A/\ln {\cal R}_L$.  However,
attempts to produce such structures in simulations of reconnection
have been disappointing (Biskamp 1984, 1986).  In numerical
simulations the X-point region  tends to collapse towards the
Sweet-Parker geometry as the Lundquist number becomes large (Biskamp
1996; Wang, Ma, \& Bhattacharjee 1996).  One way to understand this
collapse is to consider perturbations of the original X-point
geometry.  In order to maintain this geometry reconnection has to be
fast, which requires shocks in the original (Petschek) version of this
model.  These shocks are, in turn, supported by the flows driven by
fast reconnection, and fade if $L_x$ increases.  Naturally, the
dynamical range for which the existence of such shocks is possible
depends on the Lundquist number and shrinks when fluid conductivity
increases.  The apparent conclusion is that at least in the
collisional regime reconnection occurs through narrow current sheets.

In the collisionless regime the width of the current sheets may be
determined by the ion cyclotron (or Larmor) radius $r_c$ (Parker 1979)
or by the ion skin depth (Ma \& Bhattacharjee 1996; Biskamp, Schwarz,
\& Drake 1997; Shay et al.\ 1998) which differs from the former by the
ratio of $V_A$ to ion thermal velocity.  In laboratory conditions this
often leads to a current sheet thickness which is much larger than
expected (`anomalous resistivity'). However, this effect is not likely
to be important in the interstellar medium. The thickness of the
current sheet $\Delta$ scales in the Sweet-Parker scheme as
$L_x^{1/2}$. Therefore, for a sufficiently large $L_x$ the natural
Sweet-Parker sheet thickness becomes larger than the thickness
entailed by anomalous effects. Note that the ion Larmor radius $r_c$
in an interstellar magnetic field is about a hundred kilometers. One
cannot really hope to squeeze quickly the matter from many parsecs
through a slot of this size!
\begin{figure}[t]
\parbox{0.5\textwidth}{\hspace*{3em}\includegraphics[height=7cm]{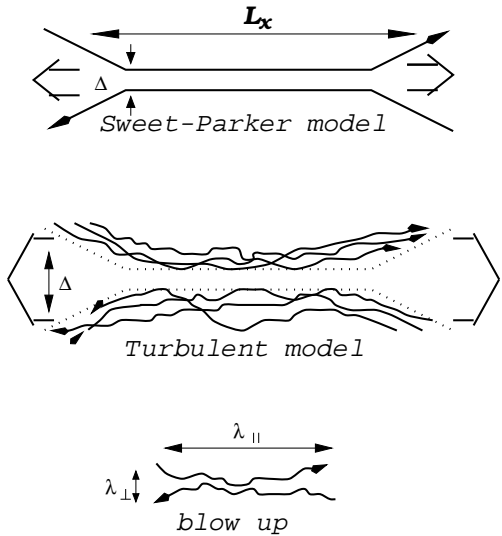}}\hfill%
\begin{minipage}{0.45\textwidth}
  \caption{({\it Top}) Sweet-Parker scheme of reconnection. ({\it
      Middle}) The new scheme of reconnection that accounts for field
    stochasticity. ({\it Bottom}) A blow up of the contact region. 
Thick arrows depict outflows of plasma.
}
\end{minipage}
\end{figure}

One may invoke anomalous resistivity to stabilize the X-point
reconnection for collisionless plasma.  For instance, Shay et al.\
(1998) found that the reconnection speed in their simulations was
independent of $L_x$, which would suggest that something like Petschek
reconnection emerges in the collisionless regime.  However, their
dynamic range was small and the ion ejection velocity increased with
$L_x$, with maximum speeds approaching $V_A$ for their largest values
of $L_x$.  Assuming that $V_A$ is an upper limit on ion ejection
speeds we may expect a qualitative change in the scaling behavior of
their simulations at slightly larger values of $L_x$. One may expect
the generic problems intrinsic to X-point reconnection to persist for
large ${\cal R}_L$.
\enlargethispage{2ex}

If neither anomalous resistivity or/and X-point reconnection work, are
there any other ways to account for fast reconnection?  Can
reconnection speeds be substantially enhanced if the plasma coupling
with magnetic field is imperfect?  This is the case in the presence of
Bohm diffusion, which is a process that is observed in laboratory
plasma but lacks a good theoretical explanation. Its characteristic
feature is that ions appear to scatter about once per Larmor
precession period. The resulting particle diffusion destroys the
`frozen-in' condition and allows significant larger magnetic field
line diffusion. The effective diffusivity of magnetic field lines is
$\eta_{\rm Bohm}\sim V_A r_c$ (see Lazarian \& Vishniac 1999,
henceforth LV99) which is a large increase over Ohmic resistivity. The
major shortcoming of this idea is that it is unclear at all whether
the concept of Bohm diffusion is applicable to astrophysical
circumstances.  Moreover, we note that even if we make this
substitution, it can produce fast reconnection, of order $V_A$, only
if $r_c\sim L_x$.  It therefore fails as an explanation for fast
reconnection for the same reason that anomalous resistivity does.

Matter may also diffuse perpendicular to magnetic field lines if the
plasma is partially ionized. Since neutrals are not directly affected
by magnetic field lines the neutral outflow layer may be much broader
than the $\Delta$ determined by Ohmic diffusivity. The trouble with
ambipolar diffusion is that ions and electrons are left in the
reconnection zone. As a result, the reconnection speed is determined
by a slow recombination process. Calculations in Vishniac \& Lazarian
(1999) show that the ambipolar reconnection rates are slow unless the
ionization ratio is extremely low.

Can plasma instabilities increase the reconnection rate?  The narrow
current sheet formed during Sweet-Parker reconnection is unstable to
tearing modes.  A study of tearing modes in LV99 showed that an
increase over the Sweet-Parker rates is possible and the resulting
reconnection rates may be as high as $V_A {\cal R}_L^{3/10}$. However,
these speeds are still exceedingly small in view of the enormous
values of ${\cal R}_L$ encountered in astrophysical plasmas.
Below we discuss a different approach to the problem of rapid
reconnection i.e., we consider magnetic reconnection\footnote{The mode
  of reconnection discussed here is sometimes is called {\it free}
  reconnection as opposed to {\it forced} reconnection. Wang et al.\ 
  (1992) define {\it free} reconnection as a process caused by a
  nonideal instability driven by the free energy stored in an
  equilibrium. If the equilibrium is stable, reconnection can be
  forced if a perturbation is applied externally.}  in the presence of
a weak random field component.

\section{Turbulent Reconnection}
\enlargethispage{2ex}

\subsection{Reconnection in Two and Three Dimensions}

Two idealizations were used in the preceding discussion. First, we
considered the process in only two dimensions.  Second, we assumed
that the magnetized plasma is laminar. The Sweet-Parker scheme can
easily be extended into three dimensions. Indeed, one can always take
a cross-section of the reconnection region such that the shared
component of the two magnetic fields is perpendicular to the
cross-section.  In terms of the mathematics nothing changes, but the
outflow velocity becomes a fraction of the total $V_A$ and the shared
component of the magnetic field will have to be ejected together with
the plasma.  This result has motivated researchers to do most of their
calculations in 2D, which has obvious advantages for both analytical
and numerical investigations.

However, physics in two and three dimensions is very different.  For
instance, in two dimensions the properties of turbulence are very
different. In LV99 we considered three dimensional reconnection in a
turbulent magnetized fluid and showed that reconnection is fast. This
result cannot be obtained by considering two dimensional turbulent
reconnection (cf.\ Matthaeus \& Lamkin 1986). Below we briefly discuss
the idea of turbulent reconnection, while the full treatment of the
problem is given in LV99.

\subsection{A Model of Turbulent Reconnection}

MHD turbulence guarantees the presence of a stochastic field component,
although its amplitude and structure clearly depends on the model we adopt
for MHD turbulence, as well as the specific environment of the field. 
We consider the case in which there exists a large scale,
well-ordered magnetic field, of the kind that is normally used as
a starting point for discussions of reconnection.  This field may,
or may not, be ordered on the largest conceivable scales.  However,
we will consider scales smaller than the typical radius of curvature
of the magnetic field lines, or alternatively, scales below the peak
in the power spectrum of the magnetic field, so that the direction
of the unperturbed magnetic field is a reasonably well defined concept.
In addition, we expect that the field has some small scale `wandering' of
the field lines.  On any given scale the typical angle by which field
lines differ from their neighbors is $\phi\ll1$, and this angle persists
for a distance along the field lines $\lambda_{\|}$ with
a correlation distance $\lambda_{\perp}$ across field lines (see Fig.~1).

The modification of the mass conservation constraint in the presence
of a stochastic magnetic field component is self-evident. Instead of
being squeezed from a layer whose width is determined by Ohmic
diffusion, the plasma may diffuse through a much broader layer,
$L_y\sim \langle y^2\rangle^{1/2}$ (see Fig.~1), determined by the
diffusion of magnetic field lines.  This suggests an upper limit on
the reconnection speed of $\sim V_A (\langle y^2\rangle^{1/2}/L_x)$.
This will be the actual speed of reconnection; the progress of
reconnection in the current sheet itself does not impose a smaller
limit. The value of $\langle y^2\rangle^{1/2}$ can be determined once
a particular model of turbulence is adopted, but it is obvious from
the very beginning that this value is determined by field wandering
rather than Ohmic diffusion as in the Sweet-Parker case.

What about limits on the speed of reconnection that arise from
considering the structure of the current sheet?  In the presence of a
stochastic field component, magnetic reconnection dissipates field
lines not over their entire length $\sim L_x$ but only over a scale
$\lambda_{\|}\ll L_x$ (see Fig.~1), which is the scale over which a
magnetic field line deviates from its original direction by the
thickness of the Ohmic diffusion layer $\lambda_{\perp}^{-1} \approx
\eta/V_{rec, local}$. If the angle $\phi$ of field deviation does not
depend on the scale, the local reconnection velocity would be $\sim
V_A \phi$ and would not depend on resistivity. In LV99 we claimed that
$\phi$ does depend on scale.  Therefore, the {\it local} reconnection
rate $V_{rec, local}$ is given by the usual Sweet-Parker formula but
with $\lambda_{\|}$ instead of $L_x$, i.e. $V_{rec, local}\approx V_A
(V_A\lambda_{\|}/\eta)^{-1/2}$.  It is obvious from Figure~1 that $\sim
L_x/\lambda_{\|}$ magnetic field lines will undergo reconnection
simultaneously (compared to a one by one line reconnection process for
the Sweet-Parker scheme). Thus, the overall reconnection rate may
be as large as $V_{rec, global}\approx V_A
(L_x/\lambda_{\|})(V_A\lambda_{\|}/\eta)^{-1/2}$.  Whether or not this
limit is important depends on the value of $\lambda_{\|}$.

The relevant values of $\lambda_{\|}$ and $\langle y^2\rangle^{1/2}$
depend on the magnetic field statistics. This calculation was
performed in LV99 using the Goldreich-Sridhar (1995) model of MHD
turbulence, the Kraichnan model (Iroshnikov 1963; Kraichnan 1965) and
for MHD turbulence with an arbitrary spectrum.  In all the cases the
upper limit on $V_{rec,global}$ was greater than $V_A$, so that the
diffusive wandering of field lines imposed the relevant limit on
reconnection speeds.  For instance, for the Goldreich-Sridhar (1995)
spectrum the upper limit on the reconnection speed was
\begin{equation}
V_{r, up}=V_A \min\left[\left({L_x\over l}\right)^{\frac{1}{2}} , 
\left({l\over L_x}\right)^{\frac{1}{2}}\right]
\left({v_l\over V_A}\right)^{2},
\label{main}
\end{equation}
where $l$ and $v_l$ are the energy injection scale and turbulent
velocity at this scale respectively.  In LV99 we also considered other
processes that can impede reconnection and find that they are less
restrictive. For instance, the tangle of reconnection field lines
crossing the current sheet will need to reconnect repeatedly before
individual flux elements can leave the current sheet behind.  The rate
at which this occurs can be estimated by assuming that it constitutes
the real bottleneck in reconnection events, and then analyzing each
flux element reconnection as part of a self-similar system of such
events.  This turns out to limit reconnection to speeds less than
$V_A$, which is obviously true regardless.  As a result, we showed in
LV99 that equation~(\ref{main}) is not only an upper limit, but is the best
estimate of the speed of reconnection.

Naturally, when turbulence is negligible, i.e. $v_l\rightarrow 0$, the
field line wandering is limited to the Sweet-Parker current sheet and
the Sweet-Parker reconnection scheme takes over. However, in practical
terms this means an artificially low level of turbulence that should
not be expected in realistic astrophysical environments.  Moreover,
the release of energy due to reconnection, at any speed, will
contribute to the turbulent cascade of energy and help drive the
reconnection speed upward.

We stress that the enhanced reconnection efficiency in turbulent
fluids is only present if 3D reconnection is considered. In this case
ohmic diffusivity fails to constrain the reconnection process as many
field lines simultaneously enter the reconnection region. The number
of lines that can do this increases with the decrease of resistivity
and this increase overcomes the slow rates of reconnection of
individual field lines. It is impossible to achieve a similar
enhancement in 2D (see Zweibel 1998) since field lines can not cross
each other.

\subsection{Energy Dissipation and its Consequences}
\enlargethispage{2ex}

It is usually believed that rapid reconnection in the limit of
vanishing resistivity implies a current singularity (Park, Monticello,
\& White 1984).  Our model does not require such singularities.
Indeed, they show that while the amount of Ohmic dissipation tends to
0 as $\eta \rightarrow 0$, the smallest scale of the magnetic field's
stochastic component decreases so that the rate of the flux
reconnection does not decrease.

The turbulent reconnection process assumes that only small segments of
magnetic field lines enter the reconnection zone and are subjected to
ohmic annihilation. Thus, only a small fraction of the magnetic energy,
proportional to ${\cal R}_L^{-2/5}$ (LV99), is released in the form of
ohmic heat. The rest of the energy is released in the form of
non-linear Alfv\'en waves that are generated as reconnected magnetic
field lines straighten up.

Naturally, the low efficiency of electron heating is of little
interest when ion and electron temperatures are tightly coupled.  When
this is not the case the LV99 model for reconnection has some
interesting consequences.  As an example, we may consider advective
accretion flows (ADAFs), following the general description given in
Narayan and Yi (1995) in which advective flows can be geometrically
thick and optically thin with a small fraction of the dissipation
going into electron heating.  If, as expected, the magnetic pressure
is comparable to the gas pressure in these systems, then a large
fraction of the orbital energy dissipation occurs through reconnection
events.  If a large fraction of this energy goes into electron heating
(cf.\ Bisnovatyi-Kogan \& Lovelace 1997) then the observational
arguments in favor of ADAFs are largely invalidated.  The results in
LV99 suggest that reconnection, by itself, will not result in
channeling more than a small fraction of the energy into electron
heating.  Of course, the fate of energy dumped into a turbulent
cascade in a collisionless magnetized plasma then becomes a critical
issue.

We also note that observations of solar flaring seem to show that
reconnection events start from some limited volume and spread as
though a chain reaction from the initial reconnection region initiated
a dramatic change in the magnetic field properties. Indeed, solar
flaring happens as if the resistivity of plasma were increasing
dramatically as plasma turbulence grows (see Dere 1996 and references
therein).  In our picture this is a consequence of the increased
stochasticity of the field lines rather than any change in the local
resistivity.  The change in magnetic field topology that follows
localized reconnection provides the energy necessary to feed a
turbulent cascade in neighboring regions.  This kind of nonlinear
feedback is also seen in the geomagnetic tail, where it has prompted
the suggestion that reconnection is mediated by some kind of nonlinear
instability built around modes with very small $k_{\|}$ (cf.\ Chang
1998 and references therein).  The most detailed exploration of
nonlinear feedback can be found in the work of Matthaeus and Lamkin
(1986), who showed that instabilities in narrow current sheets can
sustain broadband turbulence in two dimensional simulations.  Although
our model is quite different, and relies on the three dimensional
wandering of field lines to sustain fast reconnection, we note that
the concept of a self-excited disturbance does carry over and may
describe the evolution of reconnection between volumes with initially
smooth magnetic fields.

\section{Implications}

\subsection{Turbulent Reconnection and Turbulent Diffusivity}

We would like to stress that in introducing turbulent reconnection we
do not intend to revive the concept of ``turbulent diffusivity'' as
used in dynamo theories (Parker 1979).  In order to explain why
astrophysical magnetic fields do not reverse on very small scales,
researchers have usually appealed to an {\it ad hoc} diffusivity which
is many orders of magnitude greater than the ohmic diffusivity.  This
diffusivity is assumed to be roughly equal to the local turbulent
diffusion coefficient.  While superficially reasonable, this choice
implies that a dynamically significant magnetic field diffuses through
a highly conducting plasma in much the same way as a passive tracer.
 This is referred to as turbulent diffusivity and
denoted $\eta_t$, as opposed to the Ohmic diffusivity $\eta$. Its name
suggests that turbulent motions subject the field to kinematic
swirling and mixing. As the field becomes intermittent and intermixed
it can be assumed to undergo dissipation at arbitrarily high speeds.

Parker (1992) showed convincingly that the concept of turbulent
diffusion is ill-founded. He pointed out that turbulent motions are
strongly constrained by magnetic tension and large scale magnetic
fields prevent hydrodynamic motions from mixing magnetic field regions
of opposing polarity unless they are precisely anti-parallel.
However, results in LV99 show that the mobility of a magnetic field in
a turbulent fluid is indeed enhanced. For instance, due to fast
reconnection the magnetic field will not form long lasting knots.
Moreover, the magnetic field can be expected to straighten itself and
remove small scale reversals as required, in a qualitative sense, by
dynamo theory. Nevertheless, the underlying physics of this process is
very different from what is usually meant by ``turbulent
diffusivity''.  Within the turbulent diffusivity paradigm, magnetic
fields of different polarity were believed to filament and intermix on
very small scales while reconnection proceeded slowly.  On the
contrary, we have shown in LV99 that the global speed of reconnection
is fast if a moderate degree of magnetic field line wandering is
allowed.  The latter, unlike the former, corresponds to a realistic
picture of MHD turbulence and does not entail prohibitively high
magnetic field energies at small scales.

On the other hand, the diffusion of particles through a magnetized
plasma is greatly enhanced when the field is mildly stochastic.  There
is an analogy between the reconnection problem and the diffusion of
cosmic rays (Barghouty \& Jokipii 1996). In both cases charged
particles follow magnetic field lines and in both cases the wandering
of the magnetic field lines leads to efficient diffusion.

\subsection{Dynamos}

There is a general belief that magnetic dynamos operate in stars,
galaxies (Parker 1979) and accretion disks (Balbus \& Hawley
1998).  In stars, and in many accretion disks, the plasma has a high
$\beta$, that is the average plasma pressure is higher than the
average magnetic pressure. In such situations the high diffusivity of the
magnetic field can be explained by concentrating flux in
tubes\footnote{Note that flux tube formation requires initially
  high reconnection rates. Therefore, the flux tubes by themselves
  provide only a partial solution to the problem.}  (Vishniac 1995a,b).
This trick does not work in the disks of galaxies, where the magnetic
field is mostly diffuse (compare Subramanian 1998) and ambipolar
diffusion impedes the formation of flux tubes (Lazarian \& Vishniac
1996).  This is the situation where our current treatment of magnetic
reconnection is most relevant.  However, our results suggest that
magnetic reconnection proceeds regardless and that the concentration
of magnetic flux in flux tubes via turbulent pumping is not a
necessary requirement for successful dynamos in stars and accretion
discs.

To enable sustainable dynamo action and, for example, generate a
galactic magnetic field, it is necessary to reconnect and rearrange
magnetic flux on a scale similar to a galactic disc thickness within
roughly a galactic turnover time ($\sim 10^8$~years).  This implies
that reconnection must occur at a substantial fraction of the Alfv\'en
velocity.  The preceding arguments indicate that such reconnection
velocities should be attainable if we allow for a realistic magnetic
field structure, one that includes both random and regular fields.

One of the arguments against traditional mean-field dynamo theory is
that the rapid generation of small scale magnetic fields suppresses
further dynamo action (e.g., Kulsrud \& Anderson 1992). Our results
thus far show that a random magnetic field enhances reconnection by
enabling more efficient diffusion of matter from the reconnection
layer. This suggests that the existence of small scale magnetic
turbulence is a prerequisite for a successful large scale dynamo.  In
other words, we are arguing for the existence of a kind of negative
feed-back. If the magnetic field is too smooth, reconnection speeds
decrease and the field becomes more tangled.  If the field is
extremely chaotic, reconnection speeds increase, making the field
smoother. We note that it is common knowledge that magnetic
reconnection can sometimes be quick and sometimes be slow. For
instance, the existence of bundles of flux tubes of opposite polarity
in the solar convection zone indicates that reconnection can be very
slow. At the same time, solar flaring suggests very rapid reconnection
rates.

%We do not address here the controversial issue of the turbulent dynamo
%in clusters of galaxies. This was first suggested by (Jaffe 1980) and
%was elaborated in great detail by Ruzmaikin, Sokoloff \& Shukurov
%(1989), who claimed an excellent match between observations and
%predictions based on the Kazantsev (1968) theory of the turbulent
%dynamo.  However, Goldshmidt \& Rephaeli (1993) found a large ($\sim
%10^{20}$) numerical error in the value of Ohmic diffusivity used by
%Ruzmaikin et al. (1989), which formally invalidated their result.
%However, if it were possible to use the effective diffusivity
%determined by the reconnection rate instead of Ohmic diffusivity, then
%the theory of turbulent dynamo can be revived for clusters of
%galaxies.

Our results show that in the presence of MHD turbulence magnetic
reconnection is fast, and this in turn allows the possibility
of `fast' dynamos in astrophysics (see the discussion of the {\it fast
dynamo} in Parker 1992). 

Finally, we have assumed that we are dealing with a strong magnetic
field, where motions that tend to mix field lines of different
orientations are largely suppressed.  The galactic magnetic field is
usually taken to have grown via dynamo action from some extremely weak
seed field (cf.\ Zel'dovich, Ruzmaikin, \& Sokoloff 1983; Lazarian 1992
and references contained therein).  When the field is weak it can be
moved as a passive scalar and its spectrum will mimic that of
Kolmogorov turbulence.  The difference between $\lambda_{\bot}$ and
$\lambda_{\|}$ vanishes, the field becomes tangled on small scales,
and $V_{rec, local}$ becomes of the order of $V_A$.  Of course, in
this stage of evolution $V_A$ may be very small.  However, on such small
scales $V_A$ will grow to equipartition with the turbulent velocities
on the turn over time of the small eddies. The enhancement of
reconnection as $V_A$ increases accelerates the inverse cascade as
small magnetic loops merge to form larger ones.

\section{Discussion}

It is not possible to understand the dynamics of magnetized
astrophysical plasmas without understanding how magnetic fields
reconnect. Here we have compared traditional approaches to the problem
of magnetic reconnection and a new approach that includes the presence
of turbulence in the magnetized plasma.

One of the more striking aspects of our result is that the global
reconnection speed is relatively insensitive to the actual physics of
reconnection.  Equation (\ref{main}) only depends on the nature of the
turbulent cascade.  Although this conclusion was reached by invoking a
particular model for the strong turbulent cascade, we showed in LV99
that any sensible model gives qualitatively similar results.
One may say that the conclusion that reconnection is fast, even when
the local reconnection speed is slow, represents a triumph of global
geometry over the slow pace of ohmic diffusion.  In the end,
reconnection can be fast because if we consider any particular flux
element inside the contact volume, assumed to be of order $L_x^3$, the
fraction of the flux element that actually undergoes microscopic
reconnection vanishes as the resistivity goes to zero.

The new model of fast turbulent reconnection changes our understanding
of many astrophysical processes.  Firstly, it explains why
dynamos do not suppress themselves through the excessive generation of
magnetic noise, as some authors suggest (Kulsrud \& Anderson 1992).
The model also explains why reconnection may be sometimes fast and
sometimes slow, as solar activity demonstrates.  ADAFs and the
acceleration of cosmic rays at reconnection sites are other examples
of processes where a new model of reconnection should be applied.

Our results on turbulent reconnection assume that the turbulent
cascade is limited by plasma resistivity. If gas is partially ionized
collisions with neutrals may play an important role in damping
turbulence.  A study in Lazarian \& Vishniac (2000) shows that for gas
with low levels of ionization turbulent reconnection may be impeded as
magnetic field wandering is suppressed on small scales. However, the
level of suppression depends on the details of the energy injection
into the turbulent cascade (see a discussion in Lazarian \& Pogosyan
2000), which are far from being clear. Moreover, for very low
ionization levels there will be an enhancement of the reconnection
process as neutrals diffuse perpendicular to magnetic field lines.
Thus, reconnection may still be an important process in the evolution
of molecular clouds and in star formation.  

\acknowledgments
AL acknowledges valuable discussions with Chris McKee.

\end{document}